\documentclass[prd,twocolumn,
]{revtex4-1}

\usepackage{graphicx}% Include figure files
\usepackage{dcolumn}% Align table columns on decimal point

\usepackage{color} 
\usepackage{soul} 

\usepackage{bm}% bold math

\def\be{\begin{eqnarray}}
\def\ee{\end{eqnarray}}
\def\la{\langle}
\def\ra{\rangle}

\newcommand{\nn}{\nonumber\\}
\newcommand{\D}{\mathcal{D}}
\newcommand{\zrMD}{z_{\textrm{rMD}}}

\begin{document}

%%%%%%%%%%%%%%%%%%%%%%%%%%%%
\title{Self Consistent Path Sampling: Making Accurate All-Atom  Protein Folding Simulations  Possible on  Small Computer Clusters}
\author{S. Orioli}
\affiliation{ Dipartimento di Fisica,  Universit\`a degli Studi di Trento, Via Sommarive 14, Povo (Trento), I-38123 Italy}
\affiliation{INFN-TIFPA Via Sommarive 14, Povo (Trento), I-38123 Italy}
\author{S. a Beccara}
\affiliation{ Dipartimento di Fisica,  Universit\`a degli Studi di Trento, Via Sommarive 14, Povo (Trento), I-38123 Italy}
\affiliation{INFN-TIFPA Via Sommarive 14, Povo (Trento), I-38123 Italy}
\author{P. Faccioli~\footnote{Corresponding author: pietro.faccioli@unitn.it}}
\email{pietro.faccioli@unitn.it}
\affiliation{ Dipartimento di Fisica,  Universit\`a degli Studi di Trento, Via Sommarive 14, Povo (Trento), I-38123 Italy}
\affiliation{INFN-TIFPA Via Sommarive 14, Povo (Trento), I-38123 Italy}

\begin{abstract}
We introduce a powerful iterative algorithm  to compute protein folding pathways, with realistic all-atom force fields.  
Using the path integral formalism, we explicitly derive a modified Langevin equation which samples directly  the ensemble of reactive pathways,  exponentially reducing the cost of simulating thermally activated transitions. The algorithm  also yields  a rigorous stochastic estimate of the reaction coordinate.  After illustrating this approach on a simple toy model, we successfully validate it  against the results of ultra-long plain MD protein folding simulations for a fast folding protein (Fip35), which were performed on the Anton supercomputer. Using our algorithm, computing  a folding trajectory for this protein requires only $\sim 10^3$ core hours, a computational load which  could  be even carried out on a desktop workstation.  \end{abstract}
\pacs{Valid PACS appear here}
%\keywords{Suggested keywords}

\maketitle

\section{Introduction}
The protein folding pathway problem consists in clarifying the pattern of  structural changes through which a given denaturated protein reaches its native structure ~\cite{Dill review, PNAS review}. Its solution would shine light on  the main forces guiding the folding reaction and provide valuable insight on the origin of pathogenic misfolding events.

Even using the most powerful special-purpose supercomputer, plain Molecular Dynamics (MD) simulations of protein folding are feasible only for small chains (consisting of up to $\sim$ 100 amino acids),  with folding time within the ms time scale \cite{ Anton2}.    On the other hand, most proteins involved in biologically relevant folding or misfolding reactions contain several hundreds of amino-acids and have folding times which can be as long as seconds, or even minutes. 

To overcome the computational limitations of plain MD simulations,  more advanced algorithms have been proposed in literature, see e.g.  Ref.s \cite{TPS,milestoning, MSM1,metadynamics, TAMD, Tuckerman}.
Some of these techniques were successfully applied to investigate the kinetics or thermodynamics of structural reactions involving polypeptide chains, including the protein-ligand binding or even the folding of small protein fragments. However, the very slow folding reactions of complex proteins are still much beyond the reach of any of these techniques.
 
To our knowledge, the only reaction path sampling approach which has been successfully applied to characterise in full atomistic detail folding reactions of large and topologically complex proteins is the so-called Bias Functional (BF) approach \cite{BFA}.
For example,  this method was recently used to investigate folding and misfolding of several serpin proteins, which are made of nearly 400 amino-acids and have folding times as long as tens of minutes. It was shown that not only the BF method agrees with all existing experimental information on the folding mechanism, but also correctly predicts the effect of point mutations on  the protein misfolding propensity \cite{Serpinfolding}.  In Ref. \cite{PNAS2} a preliminary version of this algorithm \cite{PNAS1} was used to study a large conformational transition of the same proteins,  which occurs over about one hour. In \cite{IM7IM9} it was used to explain the puzzle of different folding kinetics  of two structurally identical proteins, while in \cite{DRPknots} it was applied to explore the folding mechanism of a protein with a knotted native state.
  
  The BF method  exploits  a  rigorous variational principle to select the most reliable folding trajectory within a set of trial  pathways,  previously generated by means of a specific type of biased dynamics, called ratchet-and-pawl MD (rMD)\cite{rMD1, rMD2}. In a rMD simulation, no bias is applied to  the protein, as long as  it spontaneously progresses towards the native state. An harmonic history-dependent force is introduced only to discourage spontaneous backtracking towards the reactant.  
  
Clearly, if this biasing force was defined in terms of a good  reaction coordinate  --for example,  the direction orthogonal to the iso-commitor hyper-surfaces in the protein configuration space--  then the rMD scheme would provide the correct description of the folding mechanism. 
In practice, however, rMD simulations of protein folding are biased along the direction set by a specific collective coordinate \cite{rMD2}, closely related to the instantaneous fraction of native contacts. Even though  the BF variational condition  is expected to improve on the results of  plain rMD simulations, a  sub-optimal  choice of biasing coordinate may give rise to systematic errors which are hard to estimate \emph{a priori}.

In this work, we introduce a reaction path sampling algorithm which enables to generate protein folding trajectories without  relying on  any model-dependent choice of biasing coordinate. Instead,  the reaction coordinate is derived self-consistently and represents  an output of the calculation, providing insight into the folding mechanism.  

This new scheme is not heuristically postulated, but rather it follows directly from the Langevin dynamics, with no additional approximation other than a mean-field estimate of some auxiliary variable. Direct comparison with the results of ultra-long plain MD simulations performed on the Anton supercomputer show that it provides a realistic representation of the reactive dynamics.  The computational cost of this new scheme is only a factor 2-3 larger than that of standard BF simulations, thus still extremely low. 

In the next section, we review the path integral representation of the Langevin dynamics,   briefly discuss the standard rMD scheme and the BF approach for protein folding simulations. Section \ref{SCPS} contains the main results of this work, providing   the mathematical derivation of our new self-consistent algorithm. In the subsequent section,  we first  illustrate its implementation on a very simple toy model  and then we apply it to perform a realistic all-atom protein folding calculations, benchmarking the results against those obtained from ultra long plain MD simulations. The main results are summarized in the conclusion section.  
  \begin{figure}[t!]\label{auxvar}
\includegraphics[width= 9 cm]{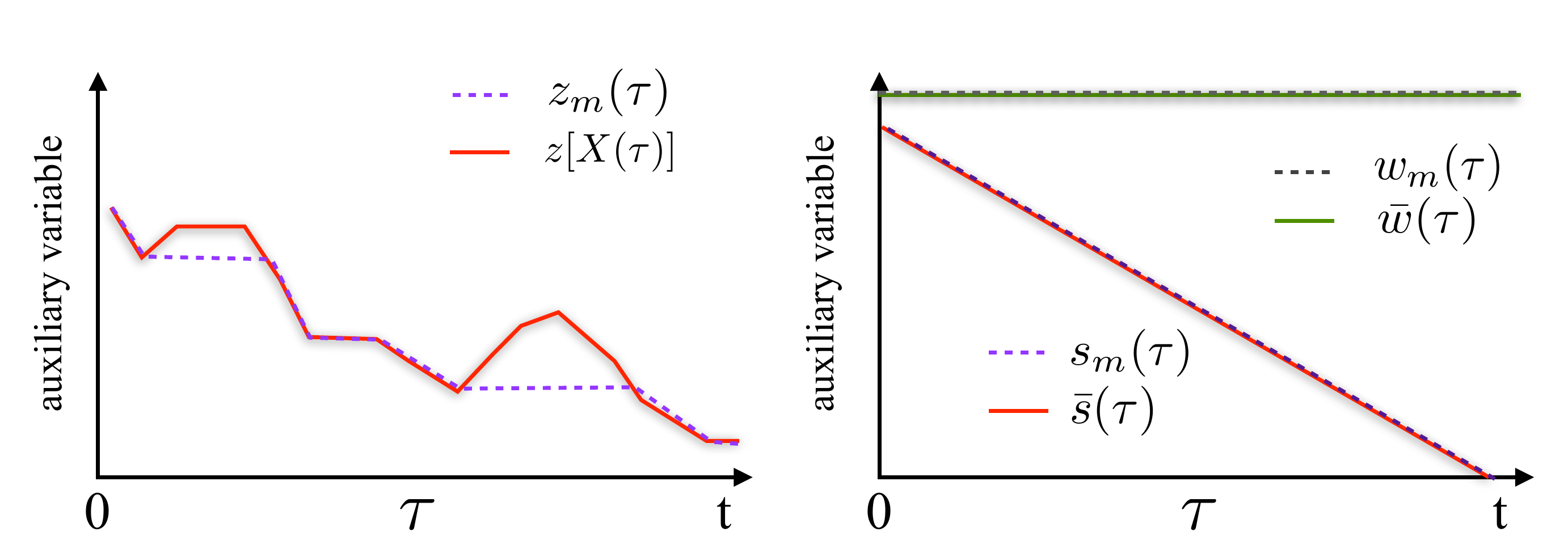}
\caption{Illustrative representation of the dynamics of the auxiliary variables introduced in the path integral representation of rMD (left panel) and in the derivation  self-consistent path sampling algorithm (right panel).  }
\end{figure}

\section{Theoretical Setup}
Throughout this paper we shall assume that the atoms in the protein obey the  Langevin equation
\be
\label{Lang}
m_i \ddot {\bf x}_i &=& -m_i \gamma_i \dot{\bf x}_i - \nabla_i U(X) +  \eta_i(t),
\ee 
where  $X=({\bf x}_1, \ldots, {\bf x}_N) $ denotes the collection of all atomic coordinates, $-\nabla_i U(X)$ is an atomistic force field, $\eta_i$ is a delta-correlated white noise obeying standard fluctuation-dissipation relationship and $m_i$ and $\gamma_i$ denote the atomic masses and viscosity, respectively. Note that, for sake of notational simplicity, we  are considering here models with an implicit solvent description. However,  all the results of the present work  hold also for an explicit solvent description, as long as the dynamics of the solvent molecules is described  by a Langevin equation. 

Within the stochastic dynamics defined by Eq.~(\ref{Lang}), the conditional probability density $p(X_N,t|X_U)$ for the protein to perform a transition from an arbitrary denatured configurations $X_U$ to a native configuration $X_N$ in a time interval $t$ can be written in path integral representation:
\be
\label{pcond}
p(X_N,t|X_U)= \int_{X_U}^{X_N}\D X~  e^{-S[X]},
\ee
where  $S[X]$ is the so-called Onsager-Machlup (OM) action:
\be\label{SOM}
&&\hspace{-0.6cm}S[X]\equiv \sum_{i=1}^N\Gamma_i\int_0^t d\tau\left(m_i \ddot{\bf x}_i + m_i \gamma_i \dot{\bf x}_i + \nabla_i U   \right)^2
\ee 
where $\Gamma_i= \frac{1}{4 \gamma_i m_i k_BT}$.

The probability for  the protein to be in the folded state at time $t$, provided it was unfolded  at the initial time  is obtained by integrating the point-to-point conditional probability (\ref{pcond}) over the final configurations in the native state and averaging over initial conditions in the unfolded state, i.e.
\be
P_{U\to N}(t) &=& \int dx_N h_N(x_N) \int d x_U h_U(x_U)\nonumber\\
&& \cdot \frac{e^{-\frac{U(X)}k_BT}}{Z} ~p(X_N ,t |X_U) 
\ee
where $h_U(X)$ and $h_N(X)$ are the characteristic functions of the unfolded and native state, respectively,  and $Z$ is the system's partition function. 
Clearly, this probability is exponentially small for time intervals $t$ much smaller than the inverse folding rate $1/k_f$. 
 
\subsection{Ratchet-and-Pawl MD and Bias Functional} \label{RMDappendix}

In order to set the stage for introducing our self-consistent path sampling algorithm, it is instructive to first review  the standard  rMD formalism and the related BF approach.
 rMD is an algorithm which enables to generate folding trajectories in time intervals $t$ much smaller the inverse folding rate. In this dynamics, an unphysical biasing force is introduced to discourage backtracking towards the unfolded state \cite{rMD1, rMD2}:  
\be\label{FrMD}
  {\bf F}_i(X,z_m) &=&  - k_R~\nabla_i z(X)~ ( z(X) - z_m)\nn
  &&\cdot~\theta(z(X) - z_m)  \ee
Here,   $z(X)$ is a collective coordinate defined as 
\be\label{zrMD}
z(X) &=& \sum_{|i-j|>35}^{N} [ C_{ij}(X)- C^0_{ij} ]^2
\ee
which measures  a Frobenius-type distance between the instantaneous contact map $C_{ij}(X)$ and the  native contact map $C_{ij}^0=  C_{ij}(X_N)$, with continuous entries defined by
\be\label{Cij}
C_{ij}(X) = \frac{1- \left(\frac{|{\bf x}_i- {\bf x}_j|}{r_0}\right)^{6}}{1- \left(\frac{|{\bf x}_i- {\bf x}_j|}{r_0}\right)^{10}}
\ee
where $r_0$~ is an arbitrary reference distance, typically set to $ 7.5$\AA.  Note that the constraint $|i-j|>35$ is introduced in order to exclude topologically closed atoms, whose relative distance is restrained by the covalent bonds. Furthermore, in order to enforce a linear scaling of the computational cost  with the number of atoms, a cut-off is usually introduced that sets to 0 the entries $C_{ij}(r_{ij})$ for atomic distances larger than a threshold, $r_{ij}>r_c$. A typical value is $r_c\simeq1.2$~nm.

 In Eq. (\ref{FrMD}) $z_m(\tau)$ is the minimum value assumed by the collective variable $z$ along the rMD trajectory, up to  time~$\tau$.  
Note that the biasing force (\ref{FrMD}) is \emph{not} active whenever the chain spontaneously evolves towards more native-like configurations ($z(t+\Delta t) < z_m$). 
It sets in only to discourage backtracking towards the unfolded state, i.e. for $z(t+\Delta t) >z_m$. 

The path integral representation of  the conditional probability to perform a transition from $X_U$ to $X_N$ in time $t$ in rMD performed within the Langevin dynamics was introduced in Ref.~\cite{BFA} and reads: 
\be\label{pcondrMD}
&&p_{\small{rMD}}(X_N, t|X_U) \equiv  \int_{X_U}^{X_N}\D X \int_{z(X_U)} \D z_m~e^{-S_{\small{rMD}}[X,z_m]}\nn
&&\cdot~ \delta\left[\dot z_m(\tau) - \dot{z}(X) ~\theta(- \dot{z}(X)) ~\theta(  z_m (\tau) - z(X) )~\right]\nn
\ee
(Throughout this paper, the Heaviside functions are conventionally defined in order to satisfy $\theta(x) =1$ for $x=0$.~)

The  expression (\ref{pcondrMD}) contains the path integral over an auxiliary time-dependent variable $z_m(\tau)$. We note that  the dynamics of  such a variable is frozen any time $z_m$  becomes smaller than $z(X)$  and any time the collective coordinate $z(X)$ is  increasing.  Its time derivative  is otherwise set equal  to $\dot z(X)$.
Therefore, by choosing  the  initial conditions $z_m(0)= z(X(0))$,  $z_m(\tau)$ is identically set equal to  the minimum value attained by the collective coordinate $z$ until time $\tau$ (see left panel of  Fig.1).

The functional $S_{\small{rMD}}[X, z_m]$ in the exponent of Eq. (\ref{pcondrMD}) coincides with an OM action with the addition of the unphysical biasing force ${\bf F}_i$:
\be
S_{\small{rMD}} &=& 
 \sum_{i=1}^N \Gamma_i \int_0^t d\tau  \left[m_i \ddot{\bf x}_i + m_i \gamma_i \dot{\bf x}_i
  + \nabla_i U  - {\bf F}_i\right]^2.\nn
\ee 
The contribution of the bias to the OM action exponentially enhances the weight of short folding pathways, making the folding probability 
$P^{\textrm{rMD}}_{U\to F}(t)\sim 1$ for time intervals $t$ much shorter than the inverse folding rate, $t\ll 1/k_F$. The prize to pay for such a computational efficiency is that of introducing uncontrolled systematic errors, which arise because the biasing coordinate $z$  may not be an optimal reaction coordinate. Furthermore, the structure of the biasing force explicitly breaks microscopic reversibility, making it impossible for rMD to directly  access thermal equilibrium. 

The systematic errors introduced by the rMD biasing scheme can be kept to a minimum  by applying the variational principle which defines the BF approach~\cite{BFA}. Indeed,  it was shown that the trajectories generated by rMD which have  the largest probability to be realized in an \emph{unbiased} Langevin simulation (i.e. for ${\bf F}^i=0$) are those with the least value of the so-called Bias Functional
\be
T[X]= \sum_{i=1}^N \frac{1}{ m_i \gamma_i} \int_0^t d\tau |{\bf F}^i[X(\tau)]|^2
\ee
Thus, in the BF approach, one generates many trial folding trajectories by rMD and then uses this variational scheme to identify the least biased pathway, which represents the variational prediction. 
\section{Self Consistent Path Sampling} 
\label{SCPS} 
Let us now introduce our new algorithm,  which provides major improvement with respect to the rMD and BF schemes discussed in the previous section. Indeed, it  follows directly from the unbiased Langevin equation and  allows us to remove the systematic errors associated to the choice of biasing coordinate.

Our starting point is path integral representation of the \emph{unbiased} Langevin dynamics (\ref{pcond}). 
We  introduce two dumb auxiliary variables $w_m(\tau)$ and $s_m(\tau)$ into this path integral by means of appropriate functional Dirac deltas:
\be\label{exact0}
&& p(X_N,t|X_U)= 
 \int_{X_U}^{X_N}\D X  \cdot e^{-S[X]}  \int_{\bar s(0)} \D s_m \int_{\bar{w}(0)} \D w_m  \nn
&& \cdot \delta\left[\dot w_m - \dot{\bar{w}} ~ \theta(-\dot{\bar{w}})~\theta(  w_m - \bar{w})~\right]  \delta\left[ \dot s_m- \dot {\bar{s}}~ \theta(-\dot{\bar{s}})~ \theta(s_m- \bar{s})\right] ,\nn
\ee
where  $\bar{s}(\tau)$ and $\bar{w}(\tau)$ are two external time-dependent functions to be defined below.
In analogy with the path integral representation of the rMD, the auxiliary variables   $s_m(\tau)$ and $w_m(\tau)$  are identically equal to the minimum value attained by  $\bar{s}$ and $\bar{w}$, until time $\tau$.  On the other hand,  we stress that  the dynamics described by the path integral (\ref{exact0}) is still unbiased, for any choice of $\bar{s}$ and $\bar{w}$. 

Let us now specialize, and define the external functions  as follows
 \be\label{sbar}
\bar s(\tau) &=& 1 - \frac{\tau}{t}\\
\label{wbar}
\bar w(\tau) &=&  w_0,
\ee
 Since $\bar{s}$ and $\bar{w}$ are never increasing,  it follows that  $s_m(\tau) = \bar{s}(\tau)$ and $w_m(\tau) = \bar{w}(\tau) $ for all times in the interval $\tau\in [0,t]$ (see right panel of Fig.1).
 
 We now observe that  Eq.s (\ref{sbar}) and (\ref{wbar}) can be equivalently written  as follows
\be\label{limit s}
\bar{s}(\tau)&=&\lim_{\lambda\to \infty} s_\lambda[X,\tau],\\
\label{limit w}
 \bar{w}(\tau) &=& \lim_{\lambda\to \infty} w_\lambda[X,\tau],
\ee
where $s_\lambda$ and $w_\lambda$ are two functionals of the path $X(\tau)$ which depend also explicitly on time $\tau$:
\be\label{sX}
\hspace{-0.5 cm}s_\lambda[X, \tau] &\equiv& \left(1- \frac{ \frac{1}{t}\int_0^t dt' ~ t' ~e^{-\lambda~||C_{ij}(\tau)- C_{ij}(t') ||^2}}{\int_0^t dt' e^{-\lambda || C_{ij}(\tau) - C_{ij}(t')||^2}}\right)\\
\label{wX}
\hspace{-0.5 cm}w_\lambda[X, \tau] &\equiv&  w_0-\frac{1}{\lambda}\log \int_0^t dt' e^{-\lambda ||C_{ij}(\tau) - C_{ij}(t')||^2}.
\ee
In these expressions $C_{ij}(\tau)$ are the  instantaneous $ij$ entry of a contact map matrix (\ref{Cij}).
 The symbol $||\ldots||$ denotes a normalized Frobenius-type distance:
%\be
%&&||C_{ij}(\tau)-C_{ij}(t')||^2\equiv \frac{\sum_{|i-j|>35}^{N} [ C_{ij}[X(\tau)]- C_{ij}[X(t')] ]^2} {N (N-1) 2}\quad
%\ee
\be
&&||C_{ij}(\tau)-C_{ij}(t')||^2\equiv \frac{\sum_{|i-j|>35}^{N} [ C_{ij}[X(\tau)]- C_{ij}[X(t')] ]^2}{\sum_{|i-j|>35}^{N} \left[ C_{ij}^0 \right]^2}\nn
\ee
where $C_0$ is the contact map calculated on the native structure of the protein.
The identities (\ref{limit s})  and (\ref{limit w}) are explicitly proven in appendix \ref{proof}, by first discretizing the time integrals in Eq.s  (\ref{sX}) and (\ref{wX})  and then noticing  that,  in the large $\lambda$ limit, the contribution  of all time slices with $t'\ne \tau$   is suppressed. 
 
  \begin{figure}[t!]\label{tubefig}
\includegraphics[width= 8 cm]{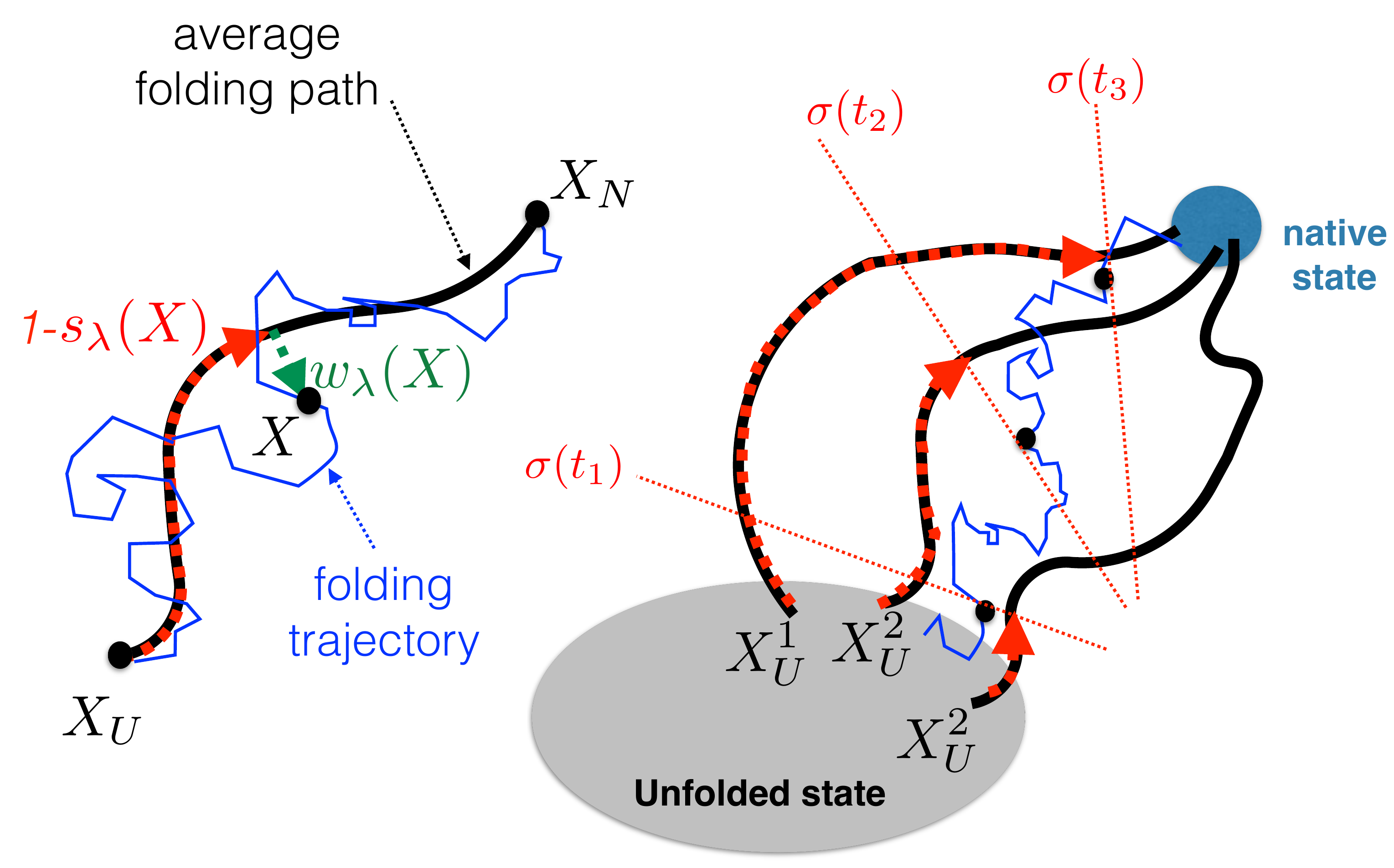}
\caption{Geometric interpretation of tube variables $s_\lambda$ and $w_\lambda$ (left panel) and of the folding reaction coordinate $\sigma$ (right panel).    }
\end{figure}

Using such equalities,  the original conditional probability density (\ref{pcond}) can be \emph{exactly} re-written as the following limit:
   \be\label{exact1}
p(X_N, t|X_U) = \lim_{\lambda\to\infty} p_\lambda(X_N, t|X_U), 
\ee
where
\be
\label{p_lambda}
&& p_\lambda(X_N t|X_U) \equiv  \int_{X_U}^{X_N}\D X \int \D s_m\int \D w_m  e^{-S_\lambda[X, s_m, w_m]} \nn
&&\delta\left[ \dot s_m- \dot s_\lambda~\theta(- \dot s_\lambda)~ \theta(  s_m-s_\lambda ) \right] ~\delta\left[\dot w_m - \dot {w}_\lambda~\theta(- \dot w_\lambda)~\right.\nn
&&\left. \theta( w_m-w_{\lambda}) \right].
 \ee
Note that the exponent in the second equation contains a new functional $S_{\lambda}[X, s_m, w_m]$. This is defined in order to maintain the structure of an OM action, but includes two additional ``forces" ${\bf F}^{w}_i$ and ${\bf F}^{s}_i$:
 \be\label{Slambda}
S_\lambda\equiv  \int_0^t d\tau\sum_{i=1}^N \Gamma_i\left[m_i \ddot{\bf x}_i + m_i \gamma_i \dot{\bf x}_i + \nabla_i  U  - {\bf F}^w_i - {\bf F}^s_i \right]^2\nn
\ee
These forces depend explicitly on $w_m$ and $s_m$, respectively and implicitly on the instantaneous configuration $X$, through  the collective variables $s_\lambda$ and $w_\lambda$:
\be\label{Fw}
&&{\bf F}^{w}_i(X, w_m) = -  k_w  \nabla_i w_\lambda~(w_\lambda -w_m)~ \theta(w_\lambda-w_m)\qquad\\
\label{Fs}
&&{\bf F}^{s}_i(X, s_m) = -  k_s  \,\nabla_i s_\lambda~(s_\lambda -s_m)~ \theta(s_\lambda-s_m)
\ee 
Their definition closely resamples that of the rMD force -- cfr Eq. (\ref{FrMD})--. However, we recall that in the large $\lambda$  limit  
\be
(w_m-w_\lambda) \to 0 \qquad \textrm{and} \qquad  (s_m -s_\lambda)\to 0.
\ee
 Thus, for sufficiently large $\lambda$, the two forces ${\bf F}_i^w$ and ${\bf F}_i^s$ are in fact always negligible, and $S_\lambda[X,s_m, w_m]$ reduces to the standard OM action  $S[X]$, proving the equivalence between Eq. (\ref{exact1}) and the original Langevin conditional probability (\ref{pcond}). 

We now introduce our \emph{only approximation} to the Langevin dynamics (\ref{Lang}): it consists in replacing the instantaneous value of the contact map $C_{ij}[X(t')]$  in the exponents in the Eq.s (\ref{sX}) and (\ref{wX}) with the average value  $\langle C_{ij}(t')\rangle$, 
\be\label{avCij}
\langle C_{ij} (t') \rangle = \frac{\int_{X_U}^{X_N} \D X~ C_{ij}[X(t')]~e^{-S[X]}}{\int_{X_U}^{X_N} \D X~e^{-S[X]}}
\ee
leading to
\be\label{asXa}
 &&\hspace{-1cm}s_\lambda[X(\tau)] \simeq 1- \frac{ \frac{1}{t}\int_0^t dt' t' e^{-\lambda~|| C_{ij}[X(\tau)] -\langle C_{ij}(t')\rangle ||^2}}{\int_0^t dt' e^{-\lambda || C_{ij}[X(\tau)] - \langle C_{ij}(t') \rangle ||^2}} \\
\label{awXa}
 &&\hspace{-1cm}w_\lambda[X(\tau)] \simeq w_0-\log \int_0^t dt' e^{-\lambda ||C_{ij}[X(\tau)] - \langle C_{ij}(t')]\rangle||^2}.
\ee
%The symbol $\langle \cdot \rangle_\lambda$  denotes an average  defined in a self-consistent way, i.e.  in terms of the dynamics defined by the path integral (\ref{p_lambda}).

Within this approximation, $w_\lambda$ and $s_\lambda$ stop depending functionally on the entire path $X(\tau)$ 
 and reduce to ordinary collective coordinates, i.e. functions of the instantaneous configuration $X(\tau)$. They represent
a specific realization  of the tube variables, introduced in Ref. \cite{tubevar} and their geometric interpretation is illustrated in Fig. 2:  $s_\lambda$ measures the progress of the reaction using as reference the self-consistently calculated average folding path, represented in contact map space.  Similarly,  $w_\lambda$ measures  in the same space the distance of a configuration  from the average folding pathway. We note that the original tube variables introduced in Ref. \cite{tubevar} were defined in terms of a \emph{fixed} external path and involved  the Euclidean distance in configuration space, instead of a distance in contact map space.  Using such a norm, however, would not enable to define a computationally viable path sampling algorithm. 

After making the mean-field replacement $C_{ij}[X(\tau)] \to \langle C_{ij}(\tau)\rangle$, the auxiliary variables $s_m$ and $w_m$ are no longer identically equal to the collective variable $s_\lambda$ and $w_\lambda$, thus the forces ${\bf F}^w_i$ and ${\bf F}^s_i$  do not vanish --see Eqs. (\ref{Fs}) and (\ref{Fw})--. As a consequence, the  path integral (\ref{exact1}) defines a new type of  rMD, with biasing forces setting in only when the collective coordinates $s_\lambda$ and $w_\lambda$ exceed their minimum value attained along the trajectory. However, unlike in the standard rMD, the biasing coordinates $s_\lambda$ and $w_\lambda$ are not arbitrarily defined \emph{a priori}. Instead, they are determined self-consistently from the reactive pathways and encode the information on  the average protein folding pathways in contact map space. In this sense, in this self-consistent type of rMD, the biasing forces act along a good reaction coordinate, thus removing the systematic uncertainties of the standard rMD. 
\begin{figure}
\begin{center}
\includegraphics[width= 9 cm]{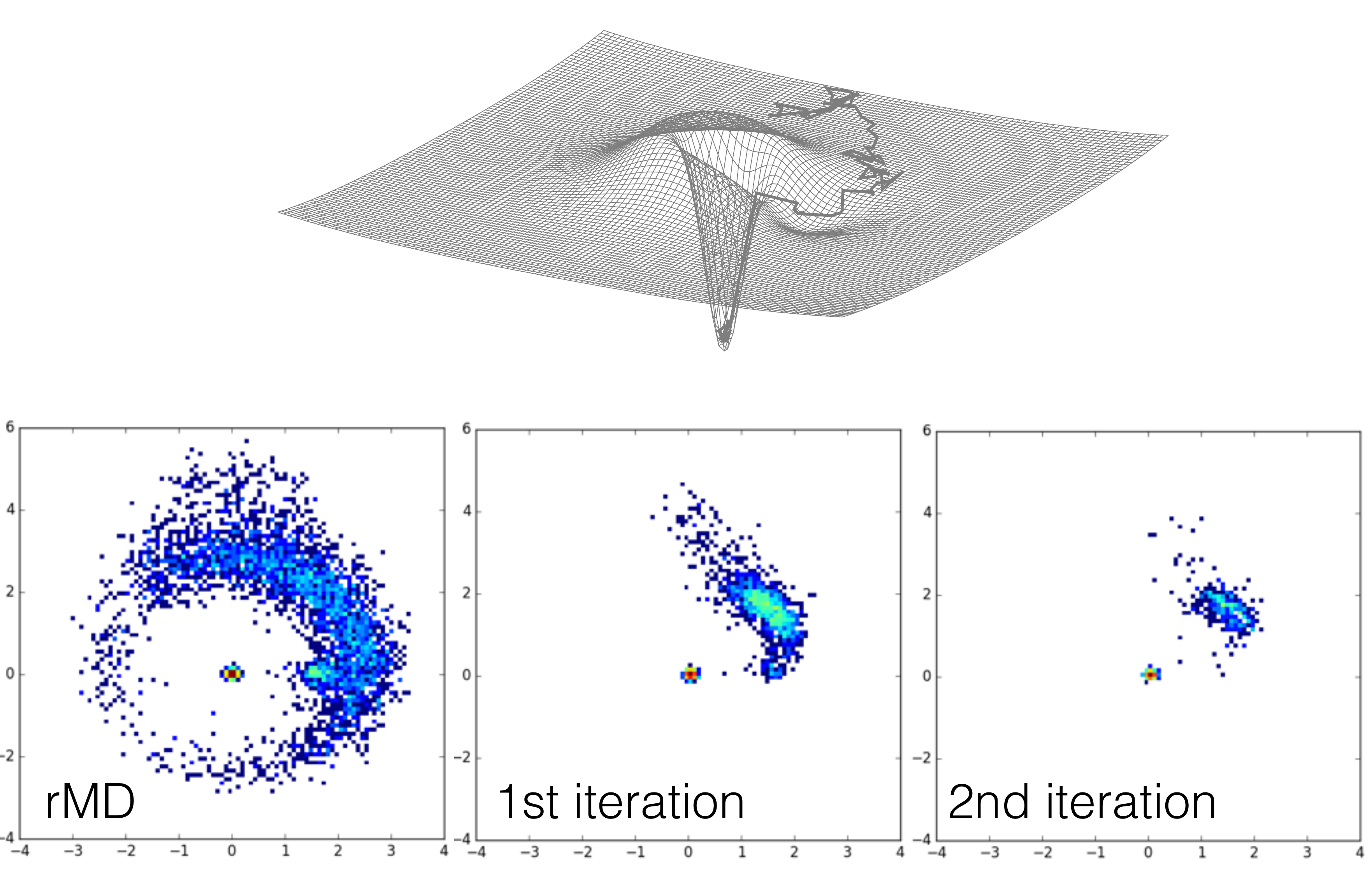}
\caption{Upper panel: two dimensional energy surface of the illustrative toy model. Lower panel:  configuration harvested by the reactive trajectories obtained at different iterations of the SCPS calculation. }
\label{S1}
\end{center}
\end{figure}

The systematic errors introduced by the mean-field  approximation $C_{ij}[X(\tau)] \to \langle C_{ij}(\tau)\rangle$ can be kept to a minimum by applying the variational principle of the BF approach: among the paths generated within this approximation, those with  largest probability to occur in the absence of any bias (i.e. after \emph{completely relaxing the mean-field approximation}) are the ones for  which the functional 
\be\label{T}
T[X]= \sum_{i=1}^N \Gamma_i\int_0^t d\tau | {\bf F}^{w}_i[X(\tau)]+{\bf F}^{s}_i[X(\tau)]|^2 
\ee
is least. 

Based on this new  approximate path integral representation of the Langevin dynamics, protein folding pathways can be sampled by means of the following iterative reaction path sampling algorithm, which we shall name Self-Consistent Path Sampling (SCPS):
\begin{enumerate}
\item An  initial denatured conditions $X_U$ is generated, for example through a thermal unfolding MD simulation, started from the native structure $X_N$;
\item By running several standard rMD simulations starting from $X_U$, an   ensemble of trial folding pathways reaching the native state within a given time interval $t$ is generated;
\item Using the trajectories evaluated in the previous step, the average contact map  $\langle C_{ij}(\tau)\rangle $ is computed for many intermediate times using Eq. (\ref{avCij}), and  the collective variables  $s_\lambda$ and $w_\lambda$ are obtained from Eq.s (\ref{sX}) and (\ref{wX}),  using a large value of $\lambda$; 
\item  A new  ensemble of trial folding pathways starting from the initial  configuration $X_U$ is obtained by performing  simulations in the new type of rMD, i.e. introducing  the biasing forces ${\bf F}_i^w$ and ${\bf F}^s_i$, based on the collective coordinates $s_\lambda$ and $w_\lambda$ evaluated at Step 3;
\item Step 3 and 4 are iterated until convergence is reached (a criterium to assess it is discussed in the next session);
\item The set of folding trajectories generated at the last iteration is scored according to the bias functional $
T[X]$ using Eq. (\ref{T})  and the least biased path is retained.
\end{enumerate}
Repeating the calculation starting from different unfolded initial conditions $X_U^1, \ldots, X_U^{N_U}$ leads to an ensemble of folding pathways. 
 
From the results of these $N_U$ independent calculations  it is possible to define a global collective coordinate $\sigma(X)$ which measures the overall progress of the folding reaction.  To this end, we combine the $N_U$ tube variables $s_\lambda^1, \ldots, s_\lambda^{N_U}$, calculated from the folding pathways started from different initial conditions (see right panel of Fig.1): 
\be\label{sigma}
\hspace{-0.6 cm} \sigma(X) \equiv \frac{1}{N_Ut}\sum_{k=1}^{N_U} \frac{\int_0^t dt' t' e^{-\lambda~|| C_{ij}[X(\tau)] -\langle C_{ij}(t')\rangle^k_\lambda ||^2}}{\int_0^t dt' e^{-\lambda || C_{ij}[X(\tau)] - \langle C_{ij}(t') \rangle^k_\lambda ||^2}}.
\ee   
In this equation, $\langle C_{ij}(\tau)\rangle_\lambda^k$ is the average contact map in the calculation started from the  initial condition $X_U^k$.

\section{Illustration and Validation}
\subsection{Diffusion in a 2-dimensional asymmetric funnel}

In order to illustrate how our self-consistent sampling scheme works, it is instructive to first apply it to a simple toy model.
In particular, we study the  diffusion of  a particle on the two-dimensional energy surface introduced in the Supplementary Material (SM) of Ref. \cite{BFA}  and defined by
\be\label{U}
U(x,y) &=& w^2 (x^2+y^2)^2  -\frac{A_1 s_1^2}{(x^2+y^2+s_1^2)^2}\nn
&&\hspace{-2cm} + \frac{A_2 s_2^2}{(x^2+y^2+s_2^2)^2} - \frac{A_3 s_3^2}{((x-x_m)^2+(y-y_m)^2+s_3^2)^2}\nn
\ee
\begin{table}[t!]
\centering
\begin{tabular}{|c  c c c c c c c c |}
\hline
 $A_1$ & $A_2$ & $A_3$ & $s_1$ & $s_2$ & $s_3$ & $w$ & $y_m$ & $x_m$ \\
 \hline
\hline
30 & 20 & 6 & 1 & 2 & 0.5 & 0.03 &0 &1.5\\ [1ex]
\hline
\end{tabular}
\caption{Parameters used in the definition of the two-dimensional asymmetric funnelled energy surface, Eq. (\ref{U})}
\end{table}%
%with $A_1=  30,      A_2=20,       A_3=6$, 
%$ s_1=1, s_2=2, s_3=0.5$,  $w=0.03$, $y_m=0$ and $x_m=1.5$.  
Using the parameters reported in Table 1, this function generates the asymmetric funnelled energy landscape shown in the upper panel of Fig 3. 

At low temperature (we choose $k_B T=0.3$) the transition across the barrier is thermally activated and the small size of the gate provides an entropic barrier.  Consequently, all reactive trajectories generated  by integrating the standard over damped Langevin equation with $\gamma \Delta t=0.02$ and initiated from an initial condition in the external ring  $(x_i=0, y_i=5)$ spend an exponentially long time before reaching the  bottom of the funnel by passing through the asymmetric gate, as shown in the upper panel of Fig.~3.

Let us now discuss the results obtained  simulating the same transition using the SCPS algorithm. We began by performing 1000 standard rMD simulations, using an harmonic biasing force with  $k_{\textrm{R}}=70$ ( see Eq. (\ref{FrMD}) )  acting along the direction set by the Euclidean distance of the particle from the center of the funnel,   $\zrMD(x,y)\equiv \sqrt{x^2+y^2}$. We emphasize that we deliberately chose to work in a worst-case scenario, i.e. we applied very a strong rMD bias ( with strength comparable to that of  the physical force)  and used  a very bad reaction coordinate, which ignores the existence of the gate through which the physical reaction pathways reach the bottom. 

The result of this rMD simulation are shown in the right panel of Fig 3. As expected, a significant fraction  of the rMD reactive trajectories reaches the bottom of the funnel by directly crossing the barrier, thus providing  a poor description of the reaction mechanism. However, the relative majority of such rMD trajectories still manages to find  the gate. As a result, the average pathway in configuration place $\langle X(\tau) \rangle$ --which in this toy model plays the role of the average contact map $\la C_{ij}(X)\ra$-- displays a small bend towards the direction of the gate. 
  \begin{figure}[t!]\label{result}
\includegraphics[width= 9 cm]{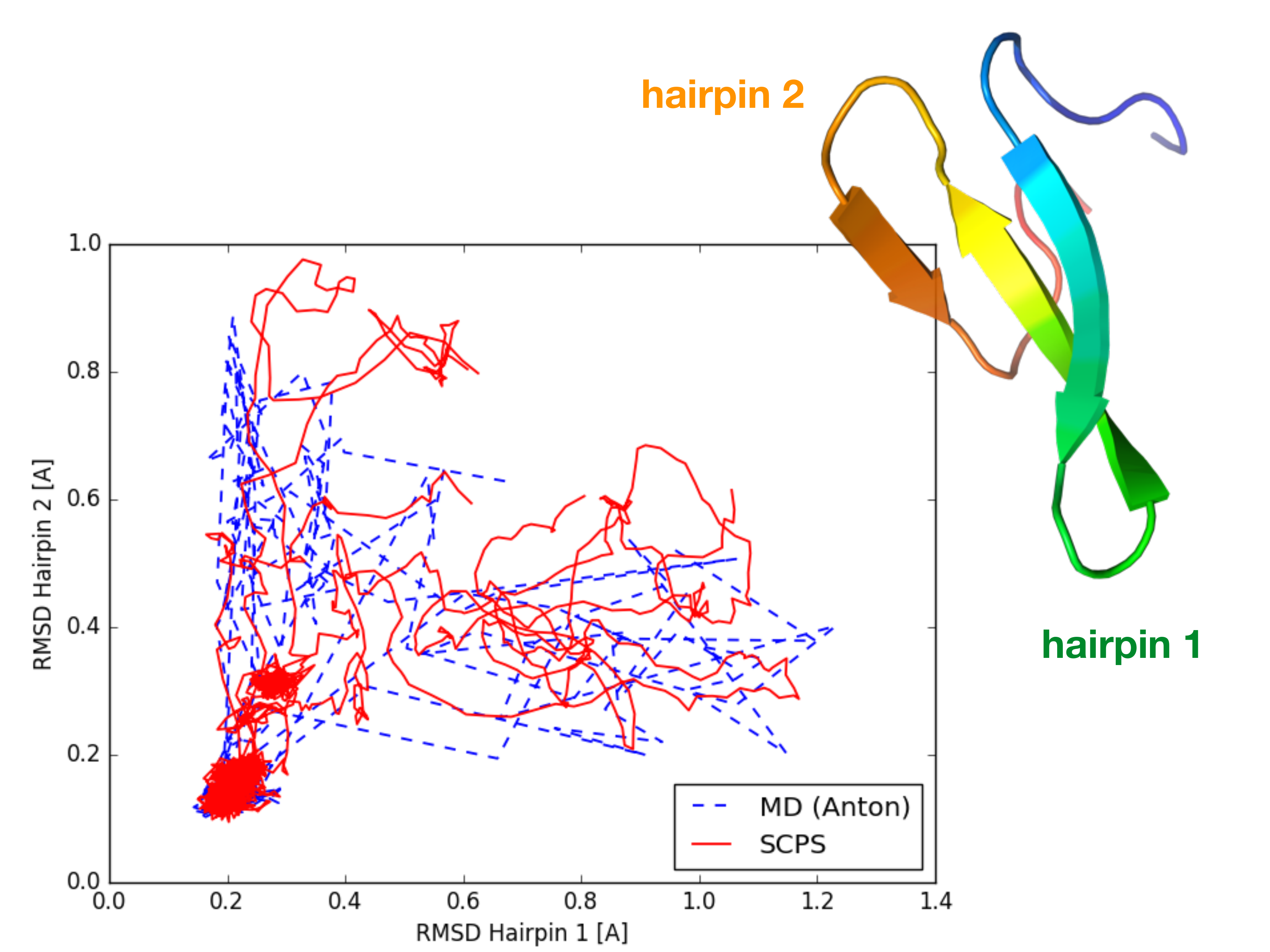}
\caption{  Comparison between folding trajectories of Fip35 (pdb: pin1,  native structure shown at the right) calculated through MD and SCPS and projected on the plane defined by the RMSD to their native structure of the two hairpins.}
\end{figure}

In all subsequent self-consistent iterations, we performed rMD simulations with the two biasing forces
\be
&&{\bf F}^{w}_i(X, w_m) = -  k_w  \nabla w_\lambda~(w_m -w_\lambda)\, \theta(w_\lambda-w_m)\nonumber\\
&&{\bf F}^{s}_i(X, s_m) = -  k_s  \,\nabla s_\lambda~(s_m -s_\lambda)\, \theta(s_\lambda-s_m), \nn
\ee 
where $k_s=3 $ and $k_w=3$ and the tube variables $w_\lambda$ and $s_\lambda$ are calculated according to
 \be
 &&\hspace{-1cm}s_\lambda(X) \simeq 1- \frac{ \frac{1}{t}\int_0^t dt' t' e^{-\lambda~|| [X(\tau) -\langle X(t')\rangle ||^2}}{\int_0^t dt' e^{-\lambda || X(\tau) - \langle X(t') \rangle||^2}} \nn
 &&\hspace{-1cm}w_\lambda(X) \simeq-\log \int_0^t dt' e^{-\lambda ||X(\tau) - \langle X(t')]\rangle||^2},
\ee
 where $||\ldots||$ denotes the Euclidean norm in configuration space.      We checked that, choosing $\lambda =0.3 $, the exponents in the definition of path variables are$\gg 1$ for most time frames.

The results shown in the lower panels of Fig 3 illustrate how, already after the first iteration, the results are significantly improved with respect to plain rMD simulations. Indeed all reactive trajectories reach the bottom by passing through the gate. The second iteration leads results consistent with the previous one, indicating that convergence has been attained.

\subsection{Realistic all-atom protein folding simulations}

Let us now assess the accuracy of the SCPS approach in a realistic protein folding simulation.  In particular, we study the folding of Fip35, the WW protein domain  shown in Fig.4, which represents a standard benchmark for protein folding simulations. Indeed, for this system, ultra-long plain MD trajectories displaying several unfolding/refolding events have been made available by DE Shaw Research \cite{Antonfip35}.

{\bf Force field:} We used the AMBER99FS-ILDN force field~\cite{Amber} with the implicit solvent model implemented in GROMACS 4.6.5 \cite{GRO4} with PLUMED 2.0.2 \cite{PLUMED}. In such an approach, the Born radii are calculated according to the Onufriev-Bashford-Case algorithm \cite{OBC}. The hydrophobic tendency of non-polar residues is taken into account through an interaction term proportional to the atomic solvent accessible surface area. The solvent-exposed surface of the different atoms is calculated from the Born radii, according to the approximation developed by Schaefer, Bartels and Karplus in \cite{BornRadii}. 
 \begin{figure}[t!]
\begin{center}
\includegraphics[width= 9 cm]{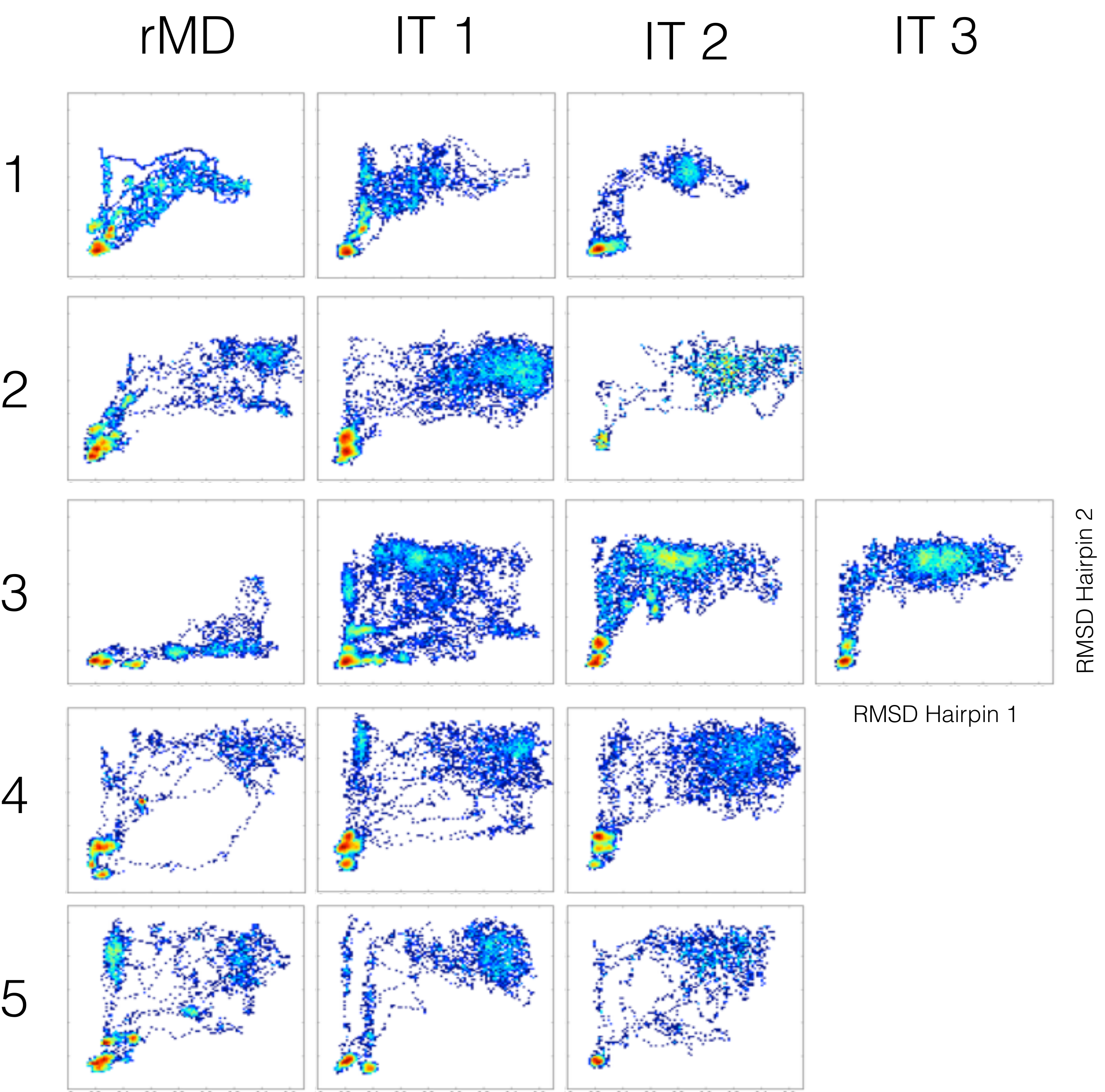}
\caption{Ensemble of folding pathways,obtained for all 5 initial conditions, at different iterations of SCPS,  projected on  the plain selected by the RMSD to native of the two hairpins.}
\label{S2}
\end{center}
\end{figure}

{\bf SCPS Implementation:} Details concerning the implementation of the steps of the SCPS algorithm introduced in the main text are given in order.
\begin{enumerate} 
\item \emph{Generation of initial conditions}: 5 independent initial conditions were generated via thermal unfolding, i.e. by running 100~ps of standard MD at T=800 K  starting from the  energy-minimized  crystal structure. 
\item \emph{Preliminary rMD simulations}: From each denatured condition, 20 independent 500-ps long  rMD trajectories were generated using a ratchet spring constant $k_R=10^{-4}$ kJ/mol. The temperature was set to T= 350 K,  which is a reasonable value for protein folding studies,  see e.g \cite{Antonfip35}. The integration time step was set $\Delta t=1$fs and  frames were saved every $0.5$ps. 
\item \emph{Self-consistent definition of collective variables}:  From each of the 5 initial conditions, the average atomic contact map  $\langle C_{ij}(\tau) \rangle$ was computed every 7 ps, using the rMD trajectories which correctly reached the folded state, defined by a Root-Mean-Square-Deviation (RMSD) to the native state less than 4~\AA.  We have checked that using  a much larger number of time frames does not significantly alter the results, yet considerably increases the computational cost of the calculation.  
 The $s_\lambda$ and $w_\lambda$ collective variables defined by 
\be
s_\lambda(X) &\simeq& 1- \frac{ \frac{1}{t}\int_0^t dt' t' e^{-\lambda~|| C_{ij}[X(\tau)] -\langle C_{ij}(t')\rangle||^2}}{\int_0^t dt' e^{-\lambda || C_{ij}[X(\tau)] - \langle C_{ij}(t') \rangle ||^2}} \\
w_\lambda(X) &\simeq&-\log \int_0^t dt' e^{-\lambda ||C_{ij}[X(\tau)] - \langle C_{ij}(t')]\rangle||^2}
\ee
were calculated from the average contact maps $\langle C_{ij}(t)\rangle$,  using  $\lambda$ = 13.5.  If much smaller values of $\lambda$ are used, then many time-frames simultaneously contribute to the time integral, signalling that the large $\lambda$ condition is not fulfilled. Conversely, much larger values of $\lambda$ lead  to lower computational efficiency. 
\item \emph{Self-consistent rMD simulations}: For each initial condition, 20 folding pathways were generated using the biasing forces: 
\be\label{Fws}
{\bf F}^{w}_i &=& -  k_w  \nabla w_\lambda~(w_\lambda -w_m)\, \theta(w_\lambda-w_m)\nonumber\\
{\bf F}^{s}_i &=& -  k_s  \,\nabla s_\lambda~(s_\lambda -s_m)\, \theta(s_\lambda-s_m)\nn
\ee 
where $k_s=2.5$ kJ/mol, $k_w=2.5\times 10^{-4}$ kJ/mol, while  $w_m(\tau)$ and $s_m(\tau)$ denote the minimum value attained by the collective coordinates $w_\lambda$ and $s_\lambda$ until time $\tau$ (see discussion in the main text).   
 \item \emph{Iteration}: Steps 3 and 4 have be repeated for two iterations in 4 out of the 5 independent simulations (corresponding to different initial conditions)  and for three iterations in the remaining simulation.  
 \item \emph{Variational correction}: For each independent simulation, we selected the minimally biased trajectory among the ensemble of folding pathways generated at the last iteration  by ranking  them according to their bias functional 
\be
T[X] = \sum_i^N \int_0^t d\tau \frac{1}{m_i \gamma_i} | {\bf F}^{w}_i + {\bf R}^s_i|^2.
\ee
\end{enumerate} 

{\bf Convergence analysis:} The density plots shown in Fig. 5 illustrate the evolution of the ensemble of  folding pathways generated at different SCPS iterations, for all 5 initial conditions.  In order to assess the convergence of the iterative algorithm, we need to  quantify how the folding pathways change from one iteration to the next. To this goal,  we have devised the following heuristic procedure.  Let  $p^{(I)}(x,y)$ be a density distribution representing how many times,  in the folding trajectories generated at the $I-$th iteration, the RMSD to native of the two hairpins assumed values $x$ and $y$ respectively.    In particular, in order to smear-out local fluctuations, we have divided  the plane identified by the RMSD to native of the two hairpins in $20\times 20$ cells of length 0.8 \AA. In the following,  the index pair $ij$ is used label the different cells, thus $p^{(I)}(x,y)\to p^{(I)}_{ij}$. 

To identify the region  visited on this plane  by the folding pathways at the $I-th$,  we consider the binary matrix:
\begin{equation}
M^{(I)}_{ij} = \left\{\begin{array}{cc}
1 & \text{ if } p^{(I)}_{ij} > 0 \\
0 &\text{ otherwise}\end{array}\right.
\end{equation}
which we further normalized to unit Frobenius norm. Finally, to compare $M_{ij}$ matrices obtained at different iterations $I$ and $J$, we computed their Frobenius distance,
\begin{equation}
D\left(M^{(I)}, M^{(J)}\right) = ||M^{(I)} - M^{(J)}||.
\end{equation}

\begin{figure}[t!]
\begin{center}
\includegraphics[width= 8 cm]{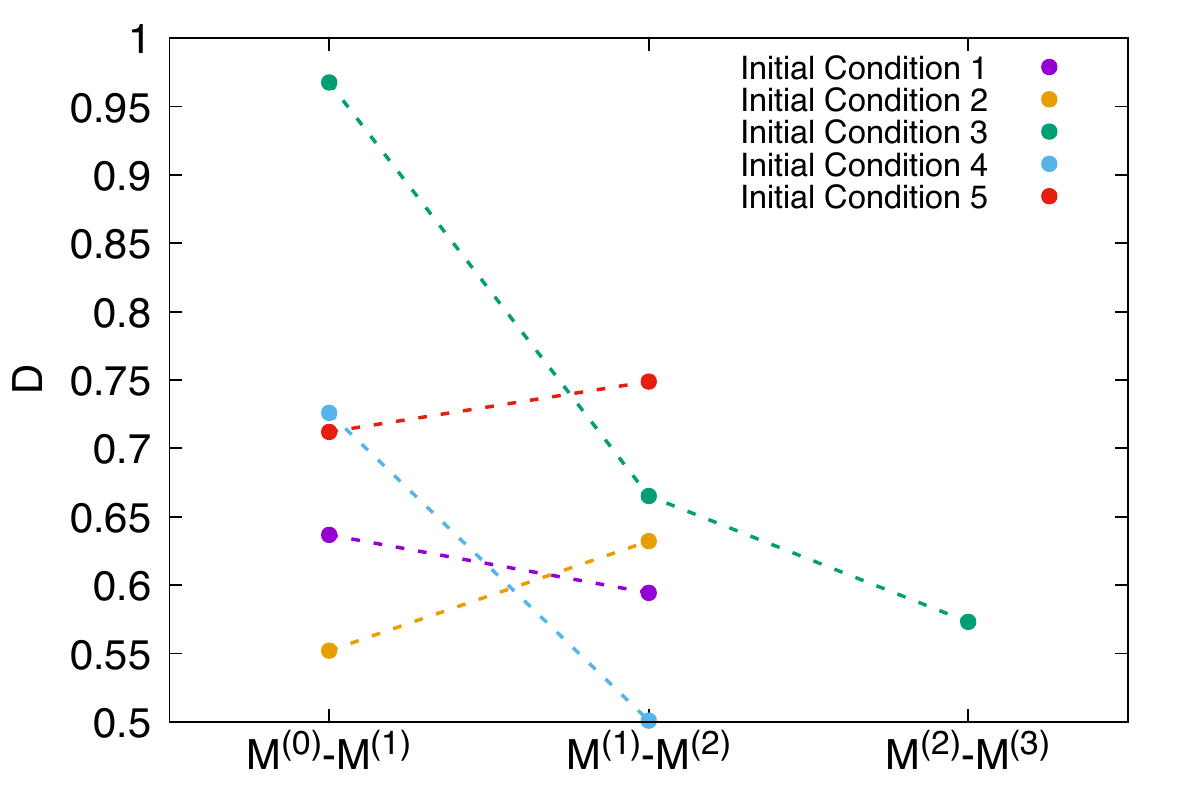}
\caption{Plot showing the approach to convergence  the SCPS simulations obtained starting from 5 different initial conditions.}
\label{S3}
\end{center}
\end{figure}

When convergence  has not yet been attained, we expect  $D\left(M^{(I)}, M^{(I+1)}\right)$ to decrease strongly when adding a new  iteration, i.e. $D\left(M^{(I)}, M^{(I+1)}\right)\gg D\left(M^{(I+1)}, M^{(I+2)}\right)$. In contrast, when convergence has been reached, we expect $D\left(M^{(I)}, M^{(I+1)}\right)\sim D\left(M^{(I+1)}, M^{(I+2)}\right) $. Note that some difference between results obtained at different iterations is expected to persist even after many iterations, due to the intrinsic stochastic character of folding trajectories.   We considered the self-consistent iterative procedure to be convergent if $D\left(M^{(I)}, M^{(I+1)}\right)$  varies less than $10\%$ from an iteration to the next. 

The plot in Fig. 6 shows the behaviour of $D\left(M^{(I)}, M^{(I+1)}\right)$ for all the $5$ initial conditions employed in the self-consistent calculation. From these results it is possible to infer that convergence has been reached in the simulations corresponding to the initial conditions 1,2,3 and 5,   while it has not yet been completely attained in the simulation associated to the initial condition 4. Note also that reaching convergence starting from the initial condition 3 required an additional iteration with respect to conditions 1,2 and 5.

{\bf Comparison with MD results:} According to the results of plain MD simulations, in the main folding pathway of this protein, hairpin-1 is completely formed before  hairpin 2 begins to fold. A less frequent alternate route is one in which the folding of the two hairpins occurs in reversed order~\cite{krivov, Pande}.  These two folding pathways are evident also in Fig. 4, which reports the folding pathways generated by MD, projected onto the plane defined by the Root-Means-Square-Deviation (RMSD) from their native structure of the two hairpins (dashed lines). The 5 folding pathways calculated with the SCPS algorithm are shown as solid lines and display  the same behavior, qualitatively demonstrating the accuracy of this algorithm, after only a few self-consistent iterations. 

In order to quantify the degree of agreement between the results of plain MD simulations and our SCPS simulations, we adopted the path similarity analysis  developed in \cite{BFA}. 
A matrix $\hat M$ is defined in order to describe the order in which the native contacts are formed \cite{rMD2}.  Namely, let $i,j$ be two indexes running over all native contacts between $C_\alpha$ atoms, and let $t_i(k)$ and $t_j(k)$ be the times  at which they are formed, i.e.
\be 
M_{ij} = \left\{
\begin{array}{cc}
1 &t_i(k)<t_j(k)\\
1/2 &t_i(k)=t_j(k)\\
0 &t_i(k)>t_j(k)
\end{array}\right.
\ee
A quantitative measure of the difference in the folding mechanisms followed by two given trajectories  $k$ and $k'$ is provided by their  path similarity $s(k,k')$, defined as
\be
s(k,k') = \frac{1}{N_c(N_c-1)}\sum_{i\ne j} \delta(M_{ij}(k)-M_{ij}(k')).
\ee
Notice that $s(k,k')=1$ if all native contacts are formed in the same order in $k$ and in $k'$, and is $0$ if they are formed in a completely different order. For comparison, we note that if $k$ and $k'$ are two random sequences of native contact formation, then $s(k, k')\sim 1/3$. 

We first computed the self-similarity  distribution, i.e. the distribution of values of $s(k,k')$, where both $k$ and $k'$ run within the ensemble of MD folding trajectories. This step is required in order to quantify the intrinsic degree of heterogeneity of the folding mechanism. Next, we computed the cross-similarity between the MD and the SCPS folding pathways, i.e. $s(k, k')$ with $k$ and $k'$ running over MD and SCPS trajectories, respectively. 
The overlap of the two distributions  shown in the Fig. 7  indicates that the average difference between the folding mechanism obtained in the two methods lies within the intrinsic statistical fluctuations. Therefore, we can conclude that the two methods give the completely consistent folding mechanisms.

It is interesting to note that the self-consistent iterations significantly improve on the  initial trial guess for reactive pathways, based  on standard rMD. Indeed,  in the first trial simulations based on rMD a significant fraction of trajectories follow a folding mechanism in which  the two hairpins form simultaneously, i.e. drawing lines close to the diagonal, in the plain defined by the RMSD to native of the two hairpins.  In the subsequent iterations, however,   the  statistical weight of such highly cooperative pathways is suppressed, leaving only pathways in which the hairpins form in order ("$L$-" and "$\Gamma-"$ shaped lines), thus improving the agreement with plain MD simulations.

\begin{figure}[t!]
\begin{center}
\includegraphics[width= 9 cm]{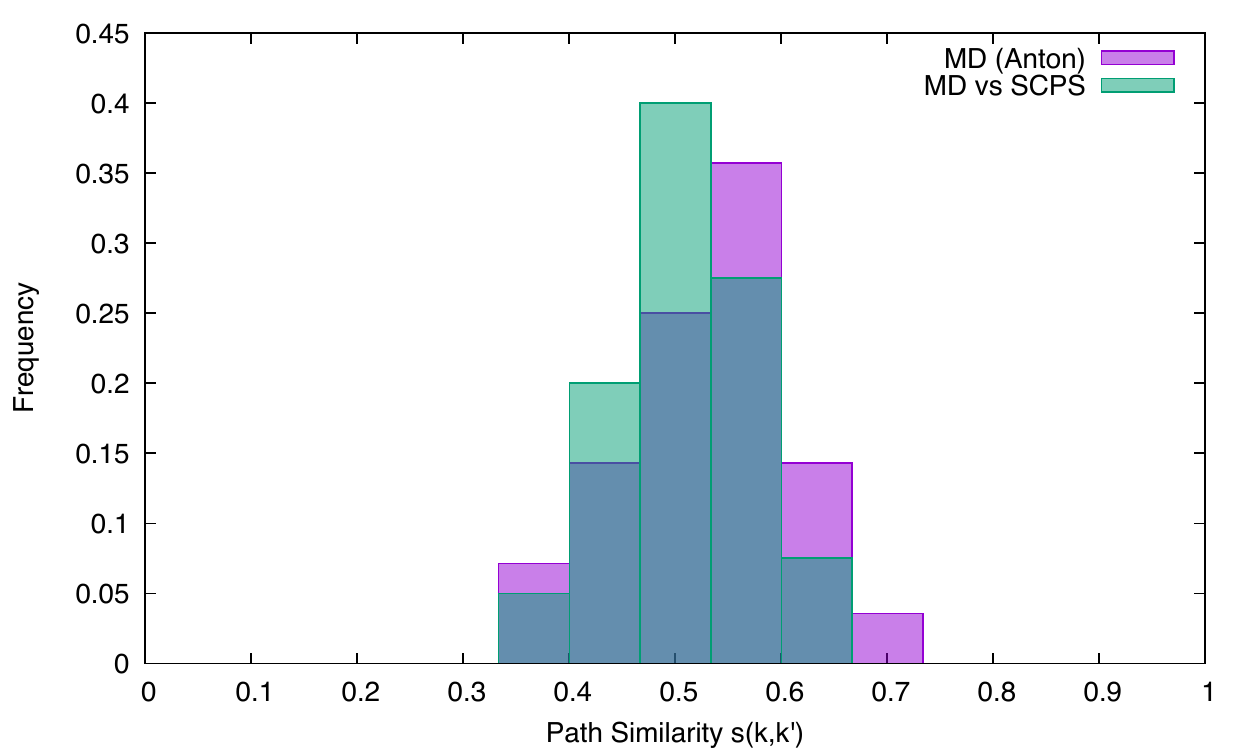}
\caption{Distribution of path similarity $s(k,k')$ for folding trajectories evaluated by plain MD and between folding trajectories evaluated in MD and SCPS.  }
\label{S4}
\end{center}
\end{figure}

\section{Conclusions}

Using the path integral representation of the Langevin dynamics, we have explicitly shown that protein folding pathways can be directly sampled through a self-consistently defined type of ratchet-and-pawl MD.

Unlike other enhanced path sampling methods, the SCPS  algorithm yields results which, at convergence, do not depend on any model-dependent choice of collective coordinate.  Instead, the algorithm  yields  a rigorous stochastic estimate of the reaction coordinate. 

 We have assessed the accuracy of  our algorithm by simulating the folding of a WW protein domain using a state-of-the-art atomistic force field, showing that it yields a folding mechanism completely consistent with the one obtained by means of  ultra-long plain MD simulations  on the Anton supercomputer. 

An important  note concerns the computational efficiency of this method. Each SCPS iteration requires to compute a few tens of very short ($< 1$ ns long) trial trajectories.  The computational cost of this scheme is therefore only a few times larger than that of the BF approach, which has been  used in the past to simulate large and complex folding processes,  with  small computer clusters. 

Finally, we emphasize that even though  in this work  we chose to focus on the prototypical protein folding pathway problem, the SCPS algorithm may be applied to a much  larger class of conformational reactions, for which structural information on the product state is experimentally available. 

\acknowledgments
The  idea of developing a self-consistent formulation of rMD arose during a joint discussion with G. Tiana and C. Camilloni, who suggested to consider tube variables. We also thank \emph{DES Research} for making available their  MD trajectories for Fip35 folding. Part of the calculations were performed on Tier-0 Marconi at the CINECA supercomputing facility.

\newpage
\appendix

\section{A Mathematical Identity}
\label{proof} 

In the following we provide the explicit proof of the identities (\ref{limit s}) and (\ref{limit w}).

We begin by discretising the time integration in Eq.s (\ref{sX}) and (\ref{wX}), with a time step $\Delta t = t/N$. Labelling the intermediate times with $\tau = l \Delta t$ we obtain
\be\label{sXapp}
s_\lambda[l \Delta t, X] &=& 1- \frac{\frac{1}{N}\sum_k~k ~e^{-\lambda~||C_{ij}[X(l \Delta t)]- C_{ij}[X(k \Delta t)] ||^2}}{\sum_k  e^{-\lambda || C_{ij}[X(l \Delta t)] - C_{ij}[X(k \Delta t)]||^2}}\nn\\
\label{wXapp}
w_\lambda[l \Delta t, X] &\equiv&  w_0-\frac{1}{\lambda}\log \Delta t\sum_k  e^{-\lambda || C_{ij}[X(l \Delta t)] - C_{ij}[X(k \Delta t)]||^2} \nn
&=& w'_0-\frac{1}{\lambda}\log \sum_k  e^{-\lambda || C_{ij}[X(l \Delta t)] - C_{ij}[X(k \Delta t)]||^2} \nn
\ee
where we redefined  the initial condition as $w'_0 = w_0 - \frac{1}{\lambda} \log \Delta t$.  
In the large $\lambda$ limit, all terms in these sums in with $l\ne k$ are suppressed, thus
\be
s_\lambda[l \Delta t, X] &\stackrel{\lambda\to \infty}{\rightarrow}& 1-\frac{l}{N} \\
w_\lambda[l \Delta t, X] &\stackrel{\lambda\to \infty}{\rightarrow}& w'_0,
\ee
After restoring the continuous notation, we recover the definitions (\ref{sbar}) and (\ref{wbar}) thus proving the identities (\ref{limit s}) and (\ref{limit w}).
\end{document}